\documentclass{article}
\usepackage{arxiv}
\usepackage{amsmath}
\usepackage[utf8]{inputenc} 
\usepackage[T1]{fontenc}  
\usepackage{url}            
\usepackage{booktabs}       
\usepackage{amsfonts}       
\usepackage{nicefrac}       
\usepackage{microtype}      
\usepackage{lipsum}
\usepackage{graphicx}
\usepackage{float}
\usepackage{multirow}
\usepackage{tabularx}
\usepackage{setspace}
\usepackage{enumerate}
\usepackage{makecell}
\usepackage{placeins}
\usepackage[colorlinks,allcolors=black]{hyperref}
\graphicspath{ {./images/} }

\title{Analyzing insurance data with an exponentiated composite Inverse-Gamma Pareto  Model}

\author{
 Bowen Liu \\
 Department of Mathematical Sciences\\
    University of Nevada, Las Vegas\\ 
    NV 89154\\

  \texttt{bowen.liu@unlv.edu} \\
   \And
 Malwane A. Ananda\\
 Department of Mathematical Sciences\\
    University of Nevada, Las Vegas\\ 
    NV 89154\\

  \texttt{malwane.ananda@unlv.edu} \\

}

\begin{document}
\maketitle
\begin{abstract}
Exponentiated models have been widely used in modeling various types of data such as survival data and insurance claims data. However, the exponentiated composite distribution models have not been explored yet. In this paper, we introduce an improvement of the one-parameter Inverse Gamma-Pareto composite model by exponentiating the random variable associated with the one-parameter Inverse Gamma-Pareto composite distribution function. The goodness-of-fit of the exponentiated Inverse Gamma-Pareto was assessed using three different insurance data sets. The two-parameter exponentiated Inverse Gamma-Pareto model outperforms the one-parameter Inverse Gamma-Pareto model in terms of goodness-of-fit measures for all datasets. In addition, the proposed exponentiated composite Inverse Gamma-Pareto model provides a very good fit with some well-known insurance datasets. 
\end{abstract}


\newpage
\section{Introduction}
\hspace{0.5cm}Modeling claim size data is one of the major topic in actuarial science. Actuaries often make decisions on financial risk management based on models. Thus, the selection of a proper model for claim sizes is a key task in the actuarial industry. Under normal circumstances, a claim size data set consists of a large number of claims with small sizes and few claims with large sizes. The common distributions in the literature such as exponential, normal, etc. do not have the ability to incorporate all the features of a claim size data set. Hence, the concept of composite distribution was introduced for modeling claim size data. With such concept, many different composite models were developed including lognormal-Pareto \cite{ananda2005,scollnik_composite_2007}, exponential-Pareto \cite{Teodorescu2006}, Weibull-Pareto \cite{preda2006}, etc. The idea of general composite model was introduced later \cite{Bak15}. With such idea, a large number of possible composite models been explored \cite{grun2019}. In general, Pareto distribution is considered good for modeling claims with large size. However, for modeling claims with small size, there are many variations in the literature. 
\par Aminzadeh and Deng introduced the  Inverse Gamma-Pareto (IG-Pareto) model recently \cite{ig_pareto} and it was suggested as a possible model for data sets with a very heavy tail such as insurance data sets. This is a one-parameter IG-Pareto composite distribution with appealing properties such as continuity and differentiability. However, fitting a one-parameter IG-Pareto model to several insurance data sets does not provide satisfactory performance, as we will show in the Numerical Examples section. Specifically, the mode of fitted IG-Pareto distribution is not large enough to describe the small claims with high frequencies within these insurance data sets. Therefore, we will modify this one-parameter IG-Pareto model by introducing an additional parameter. 

\par Exponentiated distributions were first introduced by Mudholkar and Srivastava \cite{mudholkar1990}. The main idea of exponentiated distributions is to exponentiate the Cumulative Density Function (CDF) of an existing distribution. It adds more flexibility to the traditional models due to the extra parameter. Many modifications of the existing distributions were later introduced following the idea of Mudholkar and Srivastava. For instance, Gupta and Kundu introduced exponentiated exponential \cite{gupta}; Nadarajah pioneered exponentiated beta, exponentiated Pareto and exponentiated Gamma \cite{exp_beta,exp_pareto,exp_gumbel}; Nadarajah and Gupta initiated exponentiated Gamma \cite{exp_gamma} and Afify established exponentiated Weibull-Pareto \cite{exp_weibull}. However, none of these models were established using CDF of a composite distribution. Moreover, all the exponentiated distributions mentioned above were created by exponentiating the CDF, while the exponentiated Inverse-Gamma model we propose was constructed by exponentiating the random variable associated with the CDF of a composite distribution. 

\par  The rest of the paper is organized as follows. Section 2 provides the derivation of exponentiated IG-Pareto model, the description of its behaviors and an algorithm to obtain the maximum likelihood estimators of the model. We briefly summarize the results from simulation studies in Section 3 to assess the accuracy and consistency of the MLE. In Section 4, three numerical examples are presented. Conclusions are provided in Section 5.

\section{Methodology}
\subsection{Introduction of the general composite model in loss data modeling}
Let $X$ be a positive real-valued random variable.
The general form of a composite model in loss data modeling was formally introduced \cite{scollnik_composite_2007,Bak15} as follows:

 \[ f_X(x|\alpha_1,\alpha_2,\theta,\phi) = \begin{cases} 
      \frac{1}{1+\phi}f_1^*(x|\alpha_1,\theta) & 0< x\leq \theta \\
      \frac{\phi}{1+\phi}f_2^*(x|\alpha_2,\theta) & \theta < x < \infty
   \end{cases}
\]
along with the continuity and differentiability conditions at the threshold $\theta$:
\[\begin{cases}
\text{lim}_{x \rightarrow \theta^-}f_X(x|\alpha_1,\alpha_2,\theta,\phi)= \text{lim}_{x \rightarrow \theta^+}f_X(x|\alpha_1,\alpha_2,\theta,\phi) \\
\text{lim}_{x \rightarrow \theta^-}\frac{df_X(x|\alpha_1,\alpha_2,\theta,\phi)}{dx}= \text{lim}_{x \rightarrow \theta^+}\frac{df_X(x|\alpha_1,\alpha_2,\theta,\phi)}{dx},

\end{cases}
\]
where $f_1^*$ is the probability density function of random variable $X$ when $X$ takes values between $0$ and $\theta$; $f_2^*$ is the probability density function of the random variable $X$ when $X$ takes values that are greater than $\theta$. Here, $\phi$ is a positive parameter that controls the weights of $f_1^*$ and $f_2^*$. 
\par The composite IG-Pareto model was established by Aminzadeh and Deng \cite{ig_pareto} by utilizing the theory introduced above. Suppose a random variable $X$ is known to follow a composite Inverse Gamma-Pareto distribution such that the pdf of $X$ is as follows:
\begin{equation} f_{X}(x|\theta) = \begin{cases} 
\frac{c(k\theta)^{\alpha}x^{-\alpha-1}e^{-\frac{k\theta}{x}}}{\Gamma(\alpha)} & 0 \leq x\leq \theta \\
\frac{c(\alpha-k)\theta^{\alpha-k}}{x^{\alpha-k+1}} &  x > \theta ,\\
\end{cases}
\end{equation}where, $c = 0.711384, k = 0.144351, a = 0.163947, \alpha = 0.308298$. Thus, their proposed IG-Pareto model contains only one parameter $\theta$. 
In the following subsection, we will discuss the development of exponentiated composite IG-Pareto distribution specifically. 

\subsection{Development of the exponentiated composite Inverse Gamma-Pareto distribution}

Now suppose a power transformation is applied to random variable $X$, say $Y = g(X) = X^{1/\eta}$, where $g$ is monotone increasing for any $\eta>0$. Also, $X = g^{-1}(Y) = Y^{\eta}$. For any $\eta>0$, $g^{-1}(y) = y^{\eta}$ has continuous derivative on $(0,\infty)$. Then the probability density function of $Y$ is given by: \begin{equation}
   f_{Y}(y|\theta, \eta) = \begin{cases} 
          \frac{c(k\theta)^{\alpha}(y^{\eta})^{-\alpha-1}e^{-\frac{k\theta}{y^{\eta}}}}{\Gamma(\alpha)}\eta y^{\eta-1} & 0 \leq y^{\eta}\leq \theta \\
          \frac{c(\alpha-k)\theta^{\alpha-k}}{(y^{\eta})^{\alpha-k+1}}\eta y^{\eta-1} &  y^{\eta} > \theta \\
       \end{cases} 
       \end{equation}
It can be easily shown that the above density function for exponentiated composite IG-Pareto model is continuous and differentiable on the support $(0,\infty)$.
\par The motivation for developing exponentiated IG-Pareto model as an improvement of IG-Pareto model for loss data modeling is shown in Figure 1 and 2. Two different values for $\theta$ are chosen as 5 (Figure 1) and 10 (Figure 2). For each $\theta$ value, three $\eta$ values $1,5$ and $10$ are chosen, where $\eta = 1$ corresponds to the original one-parameter IG-Pareto composite model. 
\par The figures indicate the composite exponentiated IG-Pareto model provides more flexibility to the one-parameter IG-Pareto model due to the introduction of the power parameter $\eta$. For fixed value of $\theta$, the mode of the composite exponentiated IG-Pareto increases as $\eta$ increases. 
\par In an insurance context, one of the important topics is to maximize its benefit. Given a insurance policy limit $b$ and a pdf $f_Y(y)$ associated with a loss random variable $Y$, the limited loss random variable $Y \wedge b$ is defined as following:

\[ Y \wedge b =  \begin{cases} 
Y & y \in (0,b] \\
b & y \in [b,\infty) \\
\end{cases}
\]
\par Correspondingly, the limited $t^{th}$ moment of a loss random variable $Y$, denoted by $E[(Y\wedge b)^t]$ is defined as: 
\begin{equation}
	E[(Y\wedge b)^t] = \int_{0}^{b} y^t f_Y(y) dy +\int_{b}^{\infty} b^t f_Y(y) dy
\end{equation}
Suppose $Y$ follows a exponentiated IG-Pareto distribution with parameters $\theta$ and $\eta$. It is easy to show that the $t^{th}$ limited moment of $Y$ is given by:

	\[ E[(Y\wedge b)^t] =  \begin{cases} 
	c[\frac{\Gamma(\alpha-\frac{t}{\eta},\frac{k\theta}{b^{\eta}})(k\theta)^{\frac{t} {\eta}}+b^t \Gamma(\alpha,k)-b^t \Gamma(\alpha,\frac{k\theta}{b^{\eta}})}{\Gamma(\alpha)} + b^t] & b \in (0,\theta^{1/\eta}) \\
	 c[\frac{\Gamma(\alpha-\frac{t}{\eta},k)(k\theta)^{\frac{t}{\eta}}}{\Gamma(\alpha)}+b^t]& b = \theta^{1/\eta} \\
	 c \{\frac{\Gamma(\alpha-\frac{t}{\eta},k)(k\theta)^{\frac{t}{\eta}}}{\Gamma(\alpha)}+ \frac{(\alpha-k)[b^{t-\eta(\alpha-k)}\theta^{\alpha-k}-\theta^{\frac{t}{\eta}}]}{k-\alpha+\frac{t}{\eta}}+b^{t-(\alpha-k)\eta }\theta^{\alpha-k}\}& b \in (\theta^{1/ \eta}, \infty), \\
	\end{cases}
	\]
	where $\Gamma(.,.)$ stands for an upper incomplete gamma function, $\Gamma(\alpha,x) = \int_{x}^{\infty} t^{\alpha-1}e^{-t}dt$.
\begin{figure}
	\centering
	\includegraphics[width=0.7\linewidth]{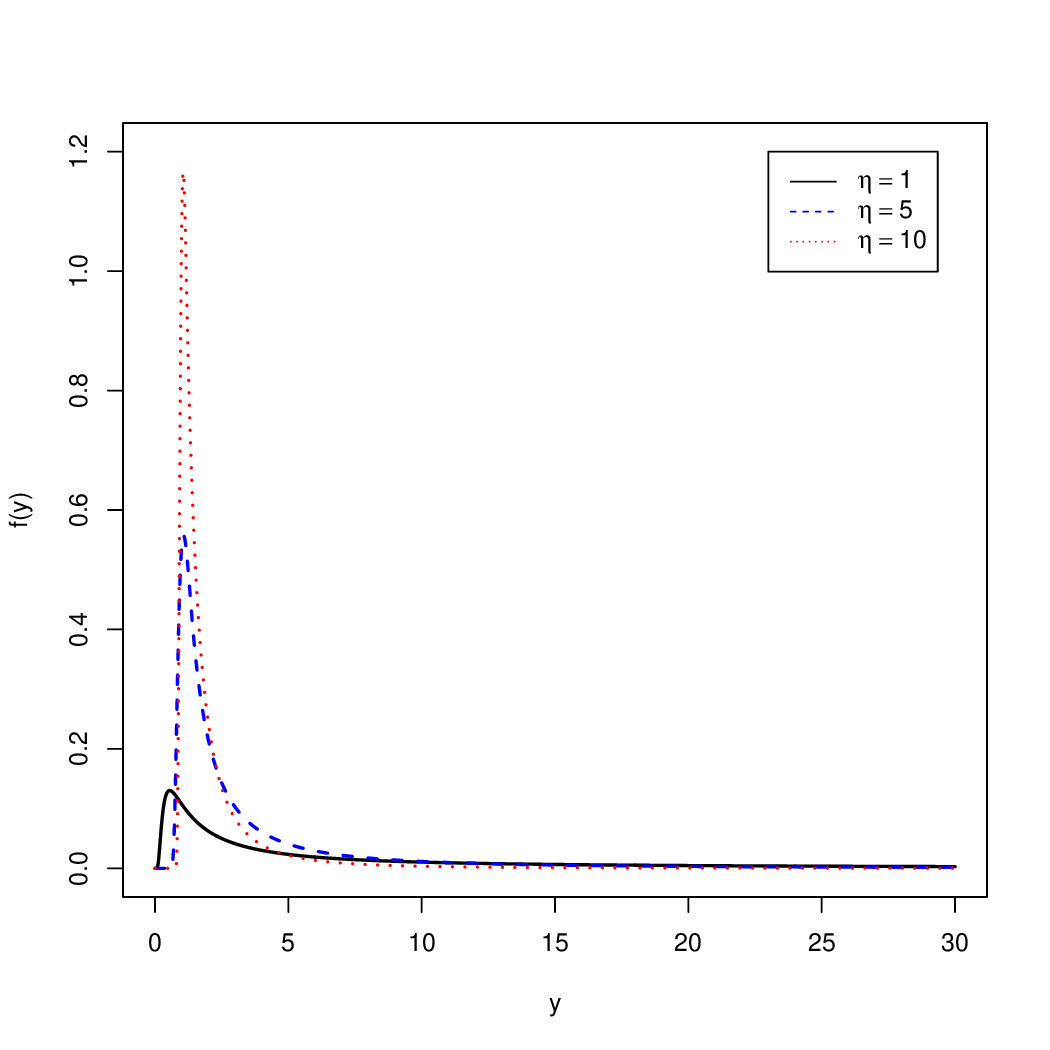}
	\caption{ Composite exponentiated IG-Pareto density ($\theta = 5$)}
	\label{fig:theta5}
\end{figure}
\begin{figure}
	\centering
	\includegraphics[width=0.7\linewidth]{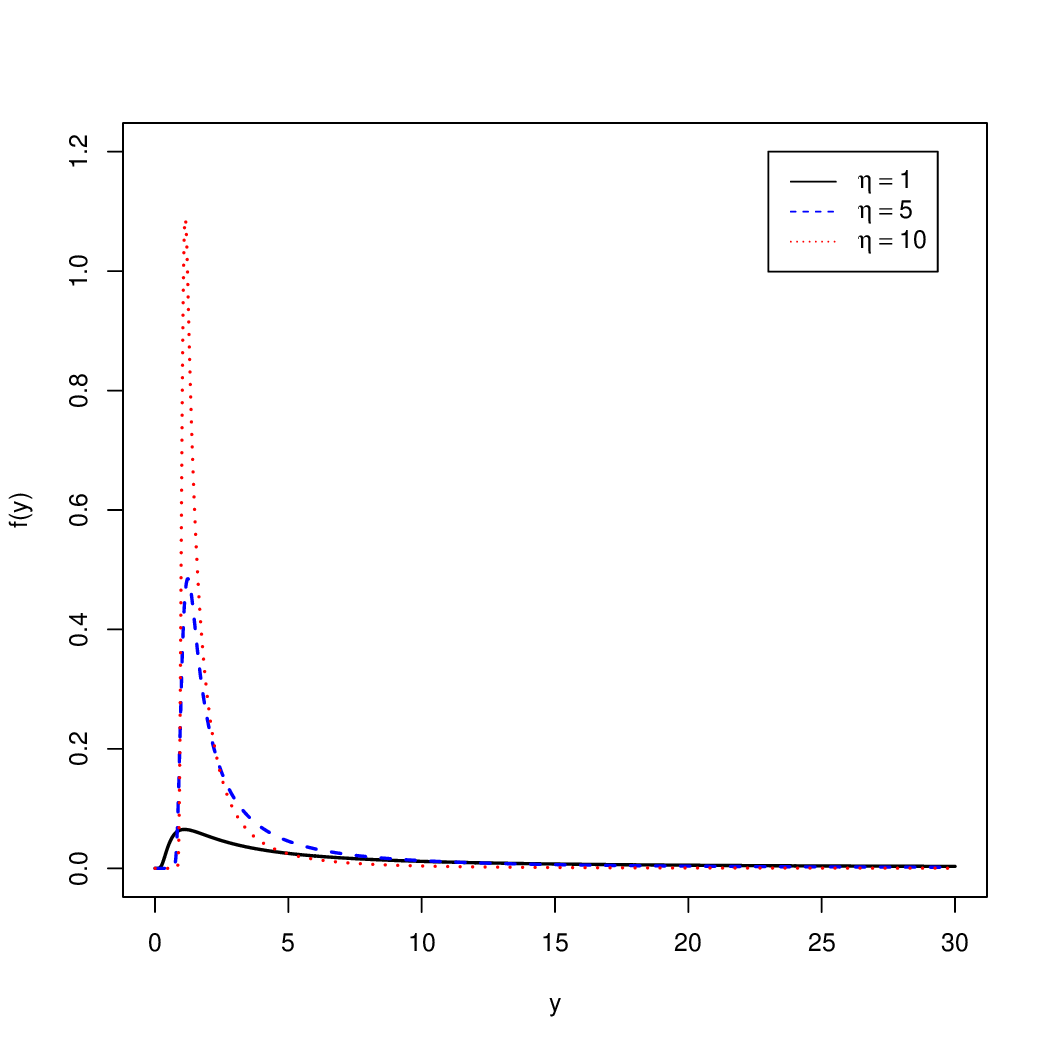}
	\caption{ Composite exponentiated IG-Pareto density ($\theta = 10$)}
	\label{fig:theta10}
\end{figure}

\pagebreak
\subsection{Parameter Estimation}
\hspace{0.5cm} Let $y_1, ... , y_n$ be a random sample from the exponentiated composite pdf given in (2). Without loss of generality, assume that $y_1 < y_2 < ... < y_n$ is an ordered random sample generated from the pdf. The likelihood function can be written as follows:
\begin{equation}
        \begin{aligned}
        L(\boldsymbol{y}|\theta, \eta) &= \prod_{i = 1}^{m}   \frac{c(k\theta)^{\alpha}(y_{i}^{\eta})^{-\alpha-1}e^{-\frac{k\theta}{y_{i}^{\eta}}}}{\Gamma(\alpha)}\eta y_i^{\eta-1}
        \prod_{j = m+1}^{n} \frac{c(\alpha-k)\theta^{\alpha-k}}{(y_{j}^{\eta})^{\alpha-k+1}} \eta y_j^{\eta-1} \\
        &= \frac{c^{n}\eta^{n}k^{\alpha m}(\alpha-k)^{n-m}(\prod_{i=1}^{m}y_i)^{-\alpha\eta-1}}{(\Gamma(\alpha))^{m}(\prod_{j=m+1}^{n}y_j)^{(\alpha-k)\eta+1}} \theta^{\alpha m + (\alpha-k) (n-m)} e^{-k\theta \sum_{i=1}^{m}\frac{1}{y_i^{\eta}}} \\
        &= Q \theta^{\alpha m + (\alpha-k) (n-m)} e^{-k\theta \sum_{i=1}^{m}\frac{1}{y_i^{\eta}}},
    \end{aligned}
    \end{equation}where
    \begin{equation*}
    Q =\frac{c^{n}\eta^{n}k^{\alpha m}(\alpha-k)^{n-m}(\prod_{i=1}^{m}y_i)^{-\alpha\eta-1}}{(\Gamma(\alpha))^{m}(\prod_{j=m+1}^{n}y_j)^{(\alpha-k)\eta+1}}.
    	\end{equation*}
    The above likelihood assumes that there exist an value $m$ such that $y_m^{\eta}<\theta<y_{m+1}^{\eta}$. The MLE of $\theta$ and $\eta$ can be obtained by solving the following equations:
  \[  \begin{cases}
        \frac{\partial L(\boldsymbol{y}|\theta, \eta)}{\partial \theta} =0\\
        \frac{\partial L(\boldsymbol{y}|\theta, \eta)}{\partial \eta} =0
    \end{cases}\]
    Closed-form expressions for MLE of $\theta$ and $\eta$ cannot be obtained. In addition, $m$ needs to be determined before finding the solution of the above equations. However, for the given values of $\eta$ and $m$, the closed-form solution of $\theta$ can be written as follows:
    \begin{equation}
        \hat{\theta}|_{\eta,m} = \frac{\alpha m +(\alpha-k)(n-m)}{k\sum_{i=1}^{m}\frac{1}{y_i^{\eta}}}
    \end{equation}
    Thus, we designed a simple search algorithm to find the MLE of $\theta$ and $\eta$ by utilizing equation (5). The description of the search algorithm is as follows: 
    \begin{enumerate}[I.]
        \item Obtain the sorted observations of a sample as $y_1 \leq y_2\leq ... \leq y_n$
        \item Determine the range of $\eta$, the parameter search will be done within the pre-defined range. Note that $\eta>0$. Hence, the search needs to be done within an left-open interval with 0 as the left endpoint. The right endpoint of this interval is data-specific. 
        \item For a known $\eta$ in the range, we start with $m=1$ and calculate the MLE of $\theta$ given $\eta$ based on (5). If $y_1^{\eta} \leq  \hat{\theta}|_{\eta,m} \leq y_2^{\eta}$, then $m = 1$. Otherwise jump to step (IV)
        \item Let $m = 2$. If $y_2^{\eta} \leq  \hat{\theta}|_{\eta,m} \leq y_3^{\eta}$, then $m = 2$. 
        We shall continue the above steps until $m$ is identified. Once $m$ is identified, keep $\hat{\theta}|_{\eta,m}$ as the MLE of $\theta$ for the known $\eta$. 
        \item Search for the optimal $\eta$ that maximizes $ L(\boldsymbol{y}|\theta,\eta)$. Find the corresponding $\hat{\theta}$ using equation (5). These are the MLEs for $\eta$ and $\theta$. 

    \end{enumerate}

\section{Simulation}
We conducted a limited simulation study to check the accuracy for the estimates of $\hat{\theta}$ and $\hat{\eta}$. For the selected sample size $n$, $\theta$ and $\eta$ values, $\mathit{N} = 5000$ samples were generated from the composite density given in (2).  
\par Table 1 to 6 present the results of all simulations under different scenarios. $\hat{\eta}_{\text{mean}}$, $\hat{\theta}_{\text{mean}}$  stand for the average of $\hat{\eta}$ and 
$\hat{\theta}$; $\hat{\eta}_{\text{SD}}$ and $\hat{\theta}_{\text{SD}}$ denote the standard deviation of $\hat{\eta}$ and $\hat{\theta}$ values, respectively. 
\par We observed that when sample size $n$ increases, the mean of the estimates of $\theta$ gets closer to the underlying true $\hat{\theta}$ under all simulation scenarios. Similarly, the mean of $\hat{\eta}$ gets closer to the underlying true $\eta$. In addition, the standard deviation of both $\hat{\theta}$ and $\hat{\eta}$ decreases as the sample size increases for different settings of the simulation parameters. Thus, the MLE of $\theta$ and $\eta$ become more accurate as the sample size increases, which is a property of maximum likelihood estimation. 
\FloatBarrier
\begin{table*}[ht]
\centering
\caption{\bf Simulation Results for $\theta = 1$ and $\eta = 0.8$}

\begin{tabularx}{\textwidth}{XXXXX}
\hline
{$n$} & $\hat{\eta}_{\text{mean}}$ & $\hat{\theta}_{\text{mean}}$& $\hat{\eta}_{\text{SD}}$& $\hat{\theta}_{\text{SD}}$
\\
\hline
20 &0.876 &	1.304	 &0.204&	1.437
\\
50 &0.828&	1.094&	0.117 &0.474
\\
100 & 0.816 &	1.040&	0.084 &	0.315
\\
500 & 0.804 &	1.006 &	0.037&0.135
\\
\hline

\end{tabularx}
\label{tab:shape-functions}
\end{table*}

\begin{table*}[ht]
	\centering
	\caption{\bf Simulation Results for $\theta = 1$ and $\eta = 1$}
	
	\begin{tabularx}{\textwidth}{XXXXX}
		\hline
		{$n$} & $\hat{\eta}_{\text{mean}}$ & $\hat{\theta}_{\text{mean}}$& $\hat{\eta}_{\text{SD}}$& $\hat{\theta}_{\text{SD}}$
		\\
		\hline
		20 &1.093	&1.262&	0.248	& 1.025
		\\
		50 &1.036 &	1.092&	0.145&	0.478
		\\
		100 & 1.017 &	1.039	&0.102&	0.307
		\\
		500 & 1.005&	1.005 &	0.049 &	0.137
		\\
		\hline
		
	\end{tabularx}
	\label{tab:shape-functions}
\end{table*}

\begin{table*}[ht]
	\centering
	\caption{\bf Simulation Results for $\theta = 1$ and $\eta = 1.2$}
	
	\begin{tabularx}{\textwidth}{XXXXX}
		\hline
		{$n$} & $\hat{\eta}_{\text{mean}}$ & $\hat{\theta}_{\text{mean}}$& $\hat{\eta}_{\text{SD}}$& $\hat{\theta}_{\text{SD}}$
		\\
		\hline
		20 &1.322	&1.263&	0.312 &	1.094
		\\
		50 &1.240 &	1.091 &	0.174 &	0.463
		\\
		100 & 1.220 &	1.048 &	0.120	& 0.314
		\\
		500 & 1.206 &	1.005 &	0.0582	& 0.140
		\\
		\hline
		
	\end{tabularx}
	\label{tab:shape-functions}
\end{table*}

\FloatBarrier

\begin{table*}[ht]
	\centering
	\caption{\bf Simulation Results for $\theta = 5$ and $\eta = 0.8$}
	
	\begin{tabularx}{\textwidth}{XXXXX}
		\hline
		{$n$} & $\hat{\eta}_{\text{mean}}$ & $\hat{\theta}_{\text{mean}}$& $\hat{\eta}_{\text{SD}}$& $\hat{\theta}_{\text{SD}}$
		\\
		\hline
		20 &0.877	&7.464&	0.203	&9.776
		\\
		50 &0.829 &	5.555 &	0.117	&1.992
		\\
		100 &0.813&	5.276&	0.082	&1.263
		\\
		500 & 0.805&	5.049 &	0.037	&0.512
		\\
		\hline
		
	\end{tabularx}
	\label{tab:shape-functions}
\end{table*}
\FloatBarrier
\begin{table*}[ht]
	\centering
	\caption{\bf Simulation Results for $\theta = 5$ and $\eta = 1.0$}
	
	\begin{tabularx}{\textwidth}{XXXXX}
		\hline
		{$n$} & $\hat{\eta}_{\text{mean}}$ & $\hat{\theta}_{\text{mean}}$& $\hat{\eta}_{\text{SD}}$& $\hat{\theta}_{\text{SD}}$
		\\
		\hline
		20 &1.098	&7.256&	0.258	& 6.480
		\\
		50 &1.036 &	5.626 &	0.146 &	2.070
		\\
		100 & 1.017 &	5.269	&0.101 &	1.232
		\\
		500 & 1.003&	5.048&	0.049&	0.511
		\\
		\hline
		
	\end{tabularx}
	\label{tab:shape-functions}
\end{table*}
\FloatBarrier
\begin{table*}[ht]
	\centering
	\caption{\bf Simulation Results for $\theta = 5$ and $\eta = 1.2$}
	
	\begin{tabularx}{\textwidth}{XXXXX}
		\hline
		{$n$} & $\hat{\eta}_{\text{mean}}$ & $\hat{\theta}_{\text{mean}}$& $\hat{\eta}_{\text{SD}}$& $\hat{\theta}_{\text{SD}}$
		\\
		\hline
		
		20 & 1.317&	7.245	&0.305&	7.213
		\\
		50 & 1.244&	5.566&	0.173	&2.025
		\\
		100 &1.224 &	5.283 &	0.121 &1.261
		\\
		500 &1.206&5.059&	0.0579	&0.511
		\\
		\hline
		
	\end{tabularx}
	\label{tab:shape-functions}
\end{table*}
\FloatBarrier

\section{Numerical Examples}
In this section, we presented the performance of the exponentiated IG-Pareto model with three different insurance data sets, namely Danish Fire Insurance Data, Norwegian Fire Insurance Data, and Society of Actuaries Group Medical Insurance Large Claims Data. In our comparisons, allong with the proposed exponentiated IG-Pareto model, we used Weibull, Inverse Gamma and IG-Pareto models. Danish Fire Insurance Data Set is well analyzed data set \cite{grun2019}. Grün and Miljkovic used 256 models to analyze the Danish Fire Insurance Data. Utilizing all those 256 models are beyond the scope of this paper. However, in their study, Weibull-Inverse Weibull came out as the best fitting model to describe the Danish Fire Insurance Data. Therefore, in addition to the abovementioned models, we included the Weibull-Inverse Weibull model in our study. Furthermore, we include the Weibull-Pareto model since we want to pick another composite model with the Pareto tail for comparison purposes. In fact, according to Grün and Miljkovic, Weibull-Pareto composite model came out as the best model among all the composite distributions with the Pareto tails.
\subsection{Goodness-of-fit of the exponentiated IG-Pareto Model}
\par To compare the performance of the different models when fitting the insurance datasets, NLL, AIC, BIC, AICc and CAIC were used. The description of the measures are listed as follows:
\begin{itemize}
	\item \textbf{NLL:} Negative Log-Likelihood is defined as the additive inverse of the loglikelihood function as follows: 
        \begin{center}
		$NLL = -logL(\boldsymbol{\hat{\theta}|y})$
	\end{center}

	\item \textbf{AIC:} Akaike's Information Criterion \cite{burnham} is defined as follows:
\begin{center}
	$AIC = -2logL(\boldsymbol{\hat{\theta}|y})+2k$,
\end{center}
where $k$ is the number of free parameters. 

	\item \textbf{BIC:} Bayesian Information Criterion \cite{burnham}  is provided as follows:  
	\begin{center}
	$BIC = -2logL(\boldsymbol{\hat{\theta}|y})+klog(n)$,
	\end{center}
where $k$ is the number of parameters and $n$ is the sample size of the data set.  
\item \textbf{AICc:} Hurvich and Tsai's Criterion \cite{aicc_1989} is provided as follows: 
\begin{center}
	$AICc = -2logL(\boldsymbol{\hat{\theta}|y})+\frac{2nk}{(n-k-1)}$,
\end{center}
\item \textbf{CAIC:} Bozdogan's criterion \cite{caic_1987} is provided as follows:
\begin{center}
$	CAIC = -2logL(\boldsymbol{\hat{\theta}|y})+k(log(n)+1)$,
\end{center}
	\end{itemize}
\par R software was used to compute the MLEs of the parameters in different models as well as NLL, AIC, BIC, AICc and CAIC values.  
\subsection{Value-at-Risk}
The estimation of extreme quantiles $Q(p)$ with $p$ being large is an important topic in insurance data modeling. These extreme quantiles are named as Value-at-Risk (VaR). For a loss random variable, VaR at the level of $p$ is defined as:
\begin{equation*}
	P(X<VaR_{p}(X)) = p,
\end{equation*}
In the context of insurance industry, VaR represents the amount of capital that an insurance company needs to have to protect the company against bankruptcy due to extreme claims.

\subsubsection{Case 1: Danish Fire Insurance Data}
\par Danish fire insurance data was widely used by many researchers to check the performance of different composite models. The data set contains 2492 claims in millions of Danish Krones (DKK) from the years 1980 to 1990. From the \textit{SMPracticals} package in R \cite{smpractical}, we were able to obtain the data and complete the analysis. The histogram of this data set is presented as Figure 3. 
\par Table 7 provides the results from this data set. Exponentiated IG-Pareto model outperforms the original one-parameter IG-Pareto model in terms of $NLL$, $AIC$ ,$BIC$, $AICc$ and $CAIC$. This is consistent with Figure 6. Firgure 6 presents the comparison of IG-Pareto model, exponentiated IG-Pareto model and the Gaussian kernel density estimate of the Danish Fire Insurance Dataset. Exponentiated IG-Pareto model provides a satisfactory fit to the Danish Fire Insurance Data while the original one-parameter IG-Pareto model does not fit the data well. Among the three two-parameter models we chose, Inverse-Gamma model performed slightly better compared to exponentiated IG-Pareto model. However, in terms of $NLL, AIC$ and $BIC$, the exponentiated IG-Pareto model gives a better performance compared to the two-parameter Weibull model. Both Weibull-Inverse Weibull and Weibull-Pareto composite models perform better than the proposed exponentiated IG-Pareto model in terms of all the goodness-of-fit measures. 
\begin{figure}
	\centering
	\includegraphics[width=0.7\linewidth]{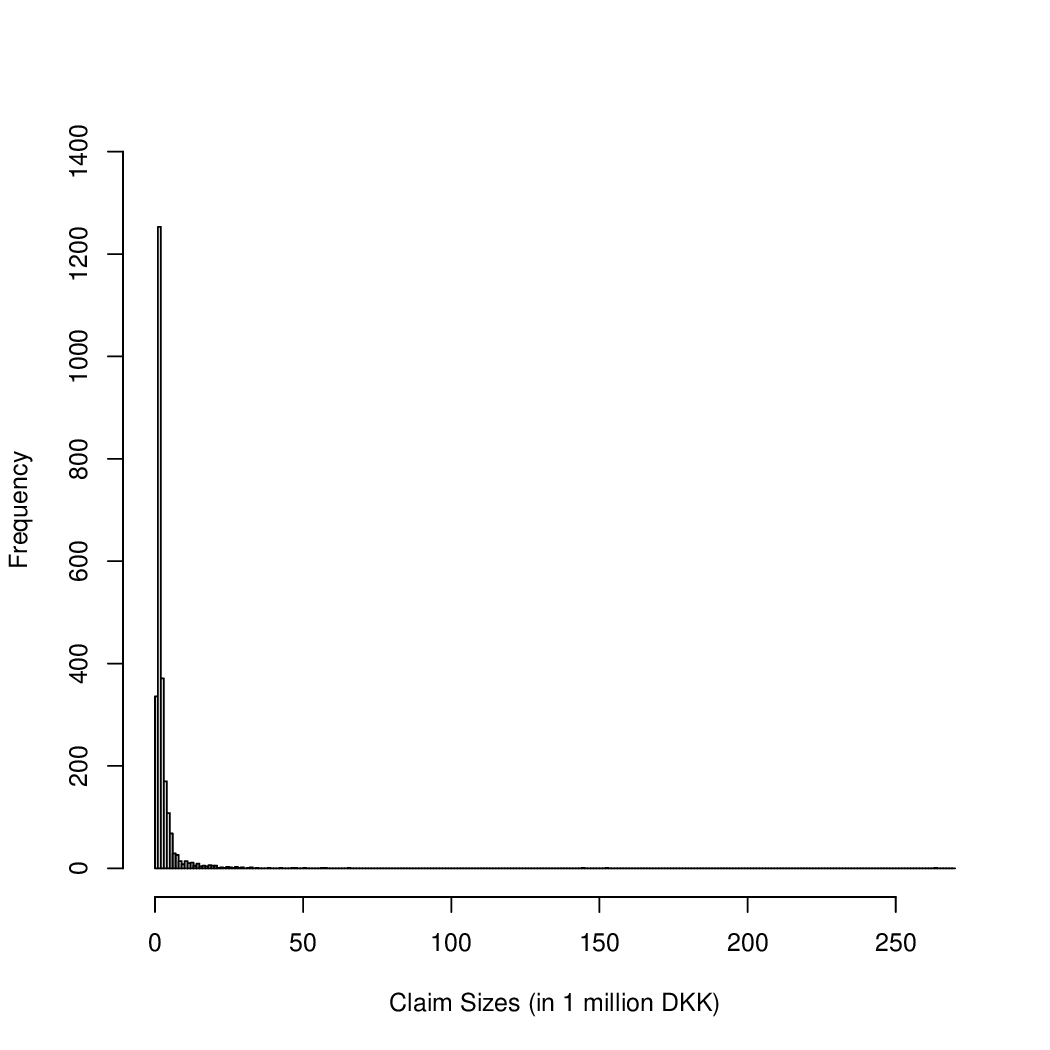}
	\caption{Histogram of Danish Fire Insurance Data Set}
	\label{fig:hist}
\end{figure}

\subsubsection{Case 2: Norwegian Fire Insurance Data}
Similar to the Danish fire insurance loss data set, the Norwegian fire insurance data was used by several researchers to investigate the performance of various loss models. The data set consists of 9181 claims in 1000s of Norwegian Krones (NKK) from the years 1972 to 1992 for a Norwegian insurance company. We obtained the data set through R package \textit{ReIns} \cite{reins}. Note the claims with size less than 500,000 NKK are forced to be 500,000 NKK. However, none of the claim values from the year 1972 are truncated, and therefore we selected the data from the year 1972 to assess the performance of the proposed model. Dealing with the truncated data is beyond the scope of this article. 
\par Figure 4 shows the histogram of the data set. The claim data from the year 1972 consists of 97 values and the claim values in millions of Norwegian Krones (NKK) are as follows:
\par 
0.520, 0.529, 0.530, 0.530, 0.544, 0.545, 0.546, 0.549, 0.553, 0.555, 0.562, 0.565, 0.565, 0.568, 0.579, 0.586, 0.600, 0.600, 0.604, 0.605, 0.621, 0.627, 0.633, 0.636, 0.667, 0.670, 0.671, 0.676, 0.681, 0.682, 0.699, 0.706, 0.725, 0.729, 0.736, 0.741, 0.744, 0.750, 0.758, 0.764, 0.767, 0.778, 0.797, 0.810, 0.849, 0.856, 0.878, 0.900, 0.916, 0.919, 0.922, 0.930, 0.942, 0.943, 0.982, 0.991, 1.051, 1.059, 1.074, 1.130, 1.148, 1.150, 1.181, 1.189, 1.218, 1.271, 1.302, 1.428, 1.438, 1.442, 1.445, 1.450, 1.498, 1.503, 1.578, 1.895, 1.912, 1.920, 2.090, 2.370, 2.470, 2.522, 2.590, 2.722, 2.737, 2.924, 3.293, 3.544, 3.961, 5.412, 5.856, 6.032, 6.493, 8.648, 8.876, 13.911, 28.055
\par Table 8 presents the goodness-of-fit results from this data set. Similar to what we observed for the Danish Fire Insurance Data, the Exponentiated IG-Pareto model performed better than the original one-parameter IG-Pareto model in terms of all goodness-of-fit measures: $NLL$, $AIC$, $BIC$, $AICc$ and $CAIC$. This is consistent with Figure 7, where exponentiated IG-Pareto model fits with the Norwegian Fire Insurance Data satisfactorily while the original one-parameter IG-Pareto model does not fit this data set well. Among the three two-parameter models we chose, exponentiated IG-Pareto model performed the best in terms of all these goodness-of-fit criteria including $NLL, AIC $ and $BIC$. In terms of $BIC$ and $CAIC$, the exponentiated IG-Pareto model also demonstrated comparable performance against the Weibull-Inverse Weibull model and the Weibull-Pareto model. However, both of these models still performed better in terms of $AIC$ and $AICc$. 
\FloatBarrier
\begin{figure}
	\centering
	\includegraphics[width=0.7\linewidth]{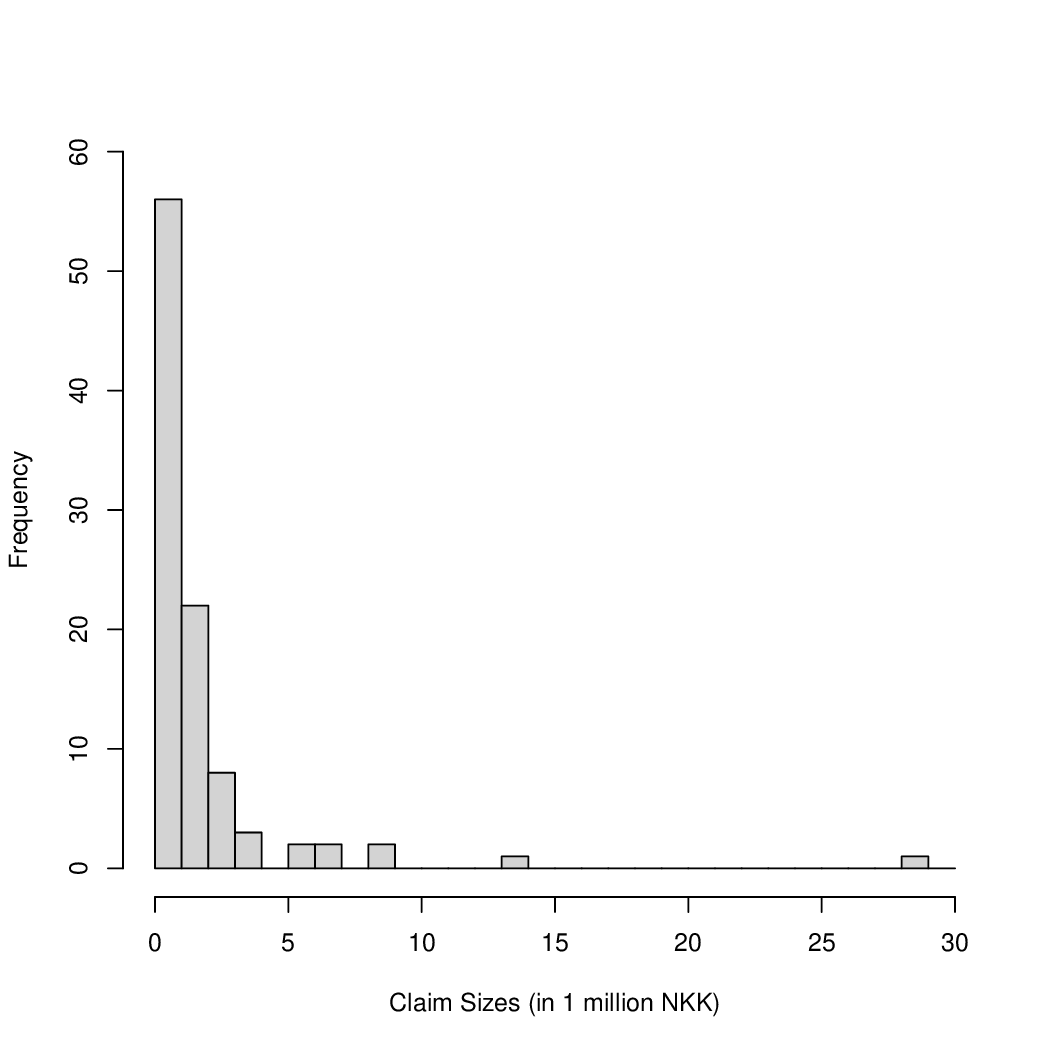}
	\caption{Histogram of Norwegian Fire Insurance Data Set (Year 1972)}
	\label{fig:hist}
\end{figure}
\FloatBarrier
\subsubsection{Case3: Society of Actuaries (SOA) Group Medical Insurance Large Claims Data}
The SOA Group Medical Insurance Claims is a publicly available data set that was published in year 1997. This data set contains 75,789 claims from year 1991 in US Dollars (USD). This data set is available in R package ReIns \cite{reins}. For the analysis concern, we rescaled the claim sizes so all the claim sizes had a unit of 10,000 USD in our analysis. The histogram of this data set is shown in Figure 5.
\par Table 9 illustrates the goodness-of-fit results for this medical insurance data set. With the medical insurance data set, exponentiated IG-Pareto model outperforms all the other methods with respect to all the goodness-of-fit measures. Figure 8 also demonstrates the exponentiated IG-Pareto model provides a very good fit to this data. Furthermore, it is clear that the one-parameter IG-Pareto model is not a good fit for this data set. 
\par Table 10 presents VaR estimates for different models at the level of $0.90, 0.95$ and $0.99$. Our model provided the closest fit compared to the empirical estimates of VaR at the level of $0.90$ and $0.95$. However, Weibull-Pareto composite model provided the closest estimate among all models at the $0.99$ level,  compared to its empirical counterparts. Notice in terms of all the goodness-of-fit measures, Weibull-Pareto is not the best model. This indicates a model that provides the best fit to the whole data set does not necessarily provides the best performance at the extreme upper tail area. 
\FloatBarrier
\begin{figure}
	\centering
	\includegraphics[width=0.7\linewidth]{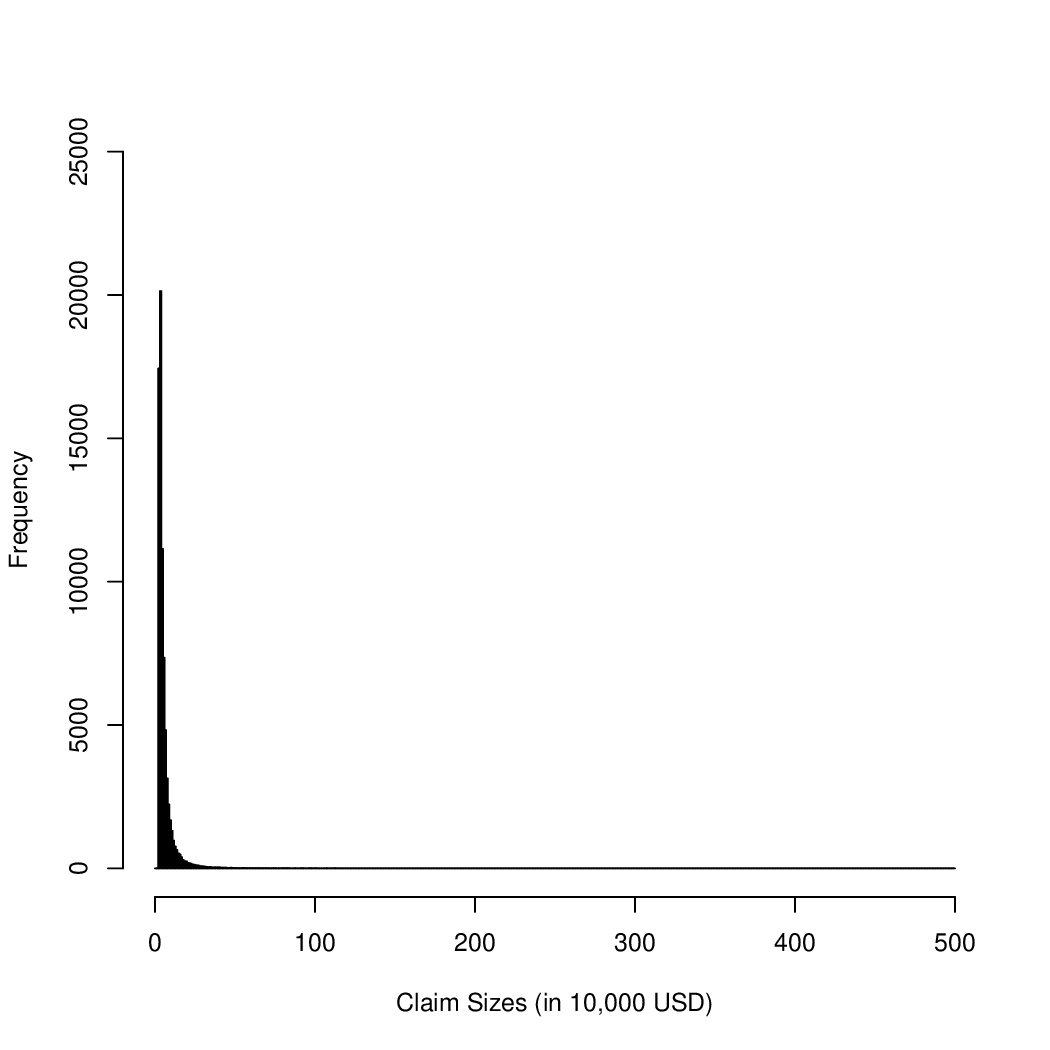}
	\caption{Histogram of SOA Group Medical Insurance Large Claims}
	\label{fig:hist}
\end{figure}
\FloatBarrier
\begin{table*}[ht]
\centering
\caption{\bf Goodness-of-fit of different models to the Danish fire data based on MLEs.}

\begin{tabular}{lllllll} 
\hline
\textbf{Model}                                                                & \textbf{p}& \textbf{NLL} & \textbf{AIC} & \textbf{BIC} & \textbf{AICc} & \textbf{CAIC}  \\ 
\hline
Weibull                                                                       & 2                                                                             & 5270.5     & 10545.0~    & 10556.6     & 10545.0      & 10558.6       \\ 
\hline
Inverse Gamma                                                                 & 2                                                                             & 4097.9     & 8199.8~    & 8211.4     & 8199.8     & 8213.4       \\ 
\hline
\begin{tabular}[c]{@{}l@{}}Inverse Gamma-Pareto\\(One-Parameter)\end{tabular} & 1                                                                             & 6983.8     & 13969.6~    & 13975.5~    & 13969.6      & 13976.5       \\ 
\hline
\begin{tabular}[c]{@{}l@{}}Exponentiated\\Inverse Gamma-Pareto\end{tabular}   & 2                                                                             & 4287.7     & 8591.0     & 8590.0     & 8579.4      & 8593.0       \\ 
\hline
Weibull-Pareto                                                                & 4                                                                             & 3823.7      & 7655.4     & 7678.6     & 7655.4      & 7682.5       \\ 
\hline
Weibull-Inverse Weibull                                                       & 4                                                                             & 3820.0      & 7648.0     & 7671.3     & 7648.0      & 7675.3       \\
\hline
\end{tabular}

  \label{tab:shape-functions}
\end{table*}

\begin{table*}[ht]
	\centering
	\caption{\bf Goodness-of-fit of different models to the Norwegian fire insurance data (year 1972) based on MLEs.}
\begin{tabular}{lllllll} 
\hline
\textbf{Model}                                                                & \textbf{p} & \textbf{NLL} & \textbf{AIC} & \textbf{BIC} & \textbf{AICc} & \textbf{CAIC}  \\ 
\hline
Weibull                                                                       & 2                                                                             & 158.7      & 321.4~    & 326.6     & 321.5      & 328.6       \\ 
\hline
Inverse Gamma                                                                 & 2                                                                             & 167.2      & 338.4     & 343.6~    & 338.6      & 345.6       \\ 
\hline
\begin{tabular}[c]{@{}l@{}}Inverse Gamma-Pareto\\(One-Parameter)\end{tabular} & 1                                                                             & 221.8      & 445.7~    & 448.2     & 445.7      & 449.2       \\ 
\hline
\begin{tabular}[c]{@{}l@{}}Exponentiated\\Inverse Gamma-Pareto\end{tabular}   & 2                                                                             & 96.1       & 196.2~    & 201.3     & 196.3      & 203.3       \\ 
\hline
Weibull-Pareto                                                                & 4                                                                             & 91.2       & 190.4~    & 200.7     & 190.8      & 204.7       \\ 
\hline
Weibull-Inverse Weibull                                                       & 4                                                                             &   90.5           &    189.0           &        199.3       &       189.4         &       203.3
         \\
\hline
\end{tabular}

	\label{tab:shape-functions}
\end{table*}

\begin{table*}[ht]
\centering
\caption{\bf Goodness-of-fit of different models to the SOA data based on MLEs.}

\begin{tabular}{lllllll} 
\hline
\textbf{Model}                                                                &  \textbf{p} & \textbf{NLL} & \textbf{AIC} & \textbf{BIC} & \textbf{AICc} & \textbf{CAIC}  \\ 
\hline
Weibull                                                                       & 2                                                                             & 204223.9     & 408451.8     & 408470.3     & 408451.8~     & 408472.3       \\ 
\hline
Inverse Gamma                                                                 & 2                                                                             & 173619.6     & 554883.8~    & 554893.0     & 554883.8      & 554894.0       \\ 
\hline
\begin{tabular}[c]{@{}l@{}}Inverse Gamma-Pareto\\(One-Parameter)\end{tabular} & 1                                                                             & 277440.9     & 347243.2~    & 347261.7     & 347243.2      & 347263.7       \\ 
\hline
\begin{tabular}[c]{@{}l@{}}Exponentiated\\Inverse Gamma-Pareto\end{tabular}   & 2                                                                             & 160836.6     & 321677.2~    & 321695.7~    & 321677.2      & 321697.7       \\ 
\hline
Weibull-Pareto                                                                & 4                                                                             & 209464.3     & 418936.6~    & 418973.5     & 418936.6~     & 418977.5       \\ 
\hline
Weibull-Inverse Weibull                                                       & 4                                                                             & 167745.8     & 335499.6     & 335536.5     & 335499.6      & 335540.5       \\
\hline
\end{tabular}
  \label{tab:shape-functions}
\end{table*}

\begin{table*}[ht]
\centering
\caption{Comparison of VaR of different models for SOA Group Medical Claims Data}
\begin{tabular}{llll} 
\hline
\textbf{Model}                                                                & \textbf{VaR(0.90)} & \textbf{VaR(0.95)} & \textbf{VaR(0.99)}  \\
\textbf{Empirical Estimates}                                                  & 10.18           & 14.76           & 30.60            \\ 
\hline
Weibull                                                                       & 12.24           & 15.01           & 20.93            \\ 
\hline
Inverse Gamma                                                                 & 9.29           & 11.78          & 19.25           \\ 
\hline
\begin{tabular}[c]{@{}l@{}}Inverse Gamma-Pareto\\(One-Parameter)\end{tabular} & 1.36 $\times 10^6$        & 9.36$\times 10^7$       & 1.72$\times 10^{12}$        \\ 
\hline
\begin{tabular}[c]{@{}l@{}}Exponentiated\\Inverse Gamma-Pareto\end{tabular}   & 9.77           & 14.54           & 36.58            \\ 
\hline
Weibull-Pareto                                                                & 13.48           & 17.66           & 27.60            \\ 
\hline
Weibull-Inverse Weibull                                                       & 8.67           & 11.40           & 21.17            \\
\hline
\end{tabular}
  \label{tab:shape-functions}
\end{table*}
\FloatBarrier
\begin{figure}
	\centering
	\includegraphics[width=0.7\linewidth]{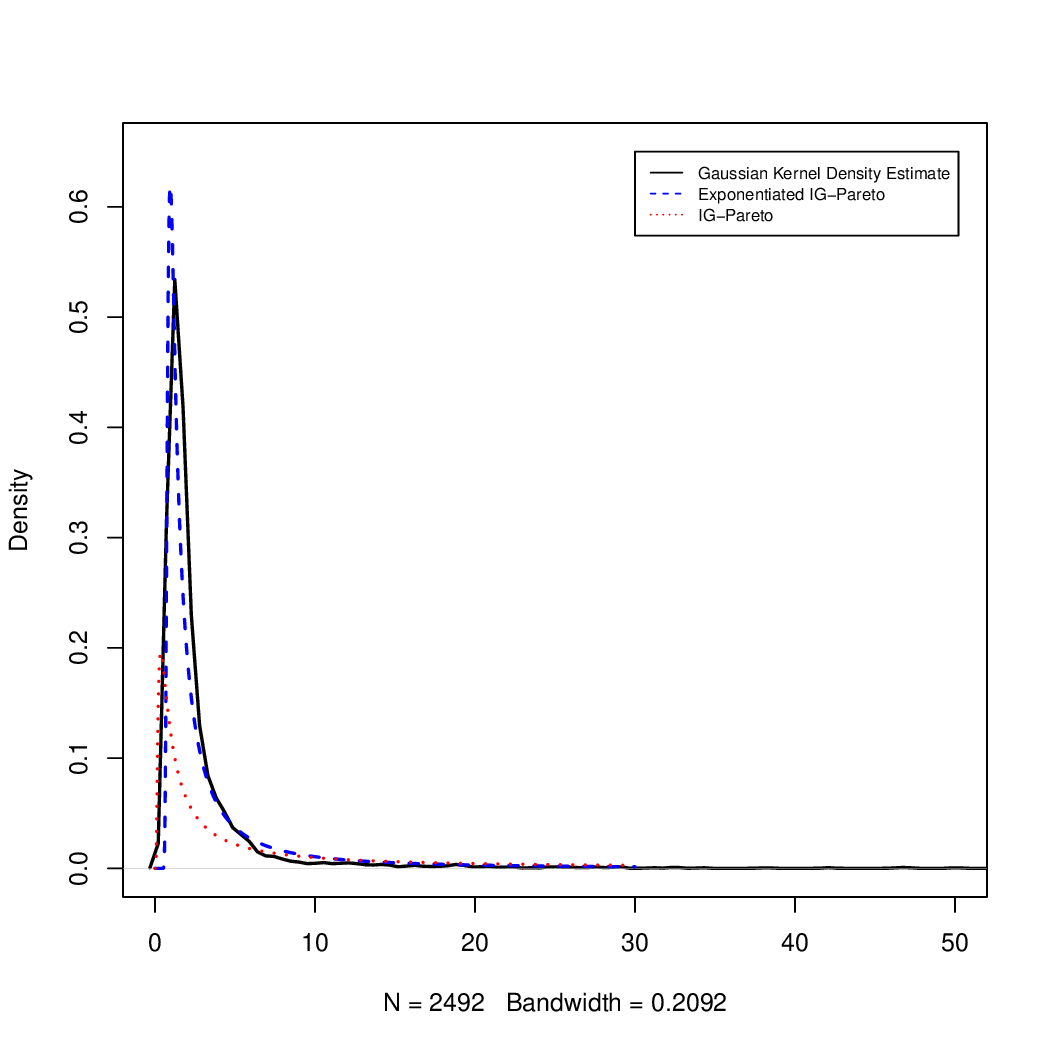}
	\caption{Density Plot of Danish Fire Insurance Data with corresponding exponentiated IG-Pareto and IG-Pareto model fit}
	\label{fig:densityplotdanish}
\end{figure}
\begin{figure}
	\centering
	\includegraphics[width=0.7\linewidth]{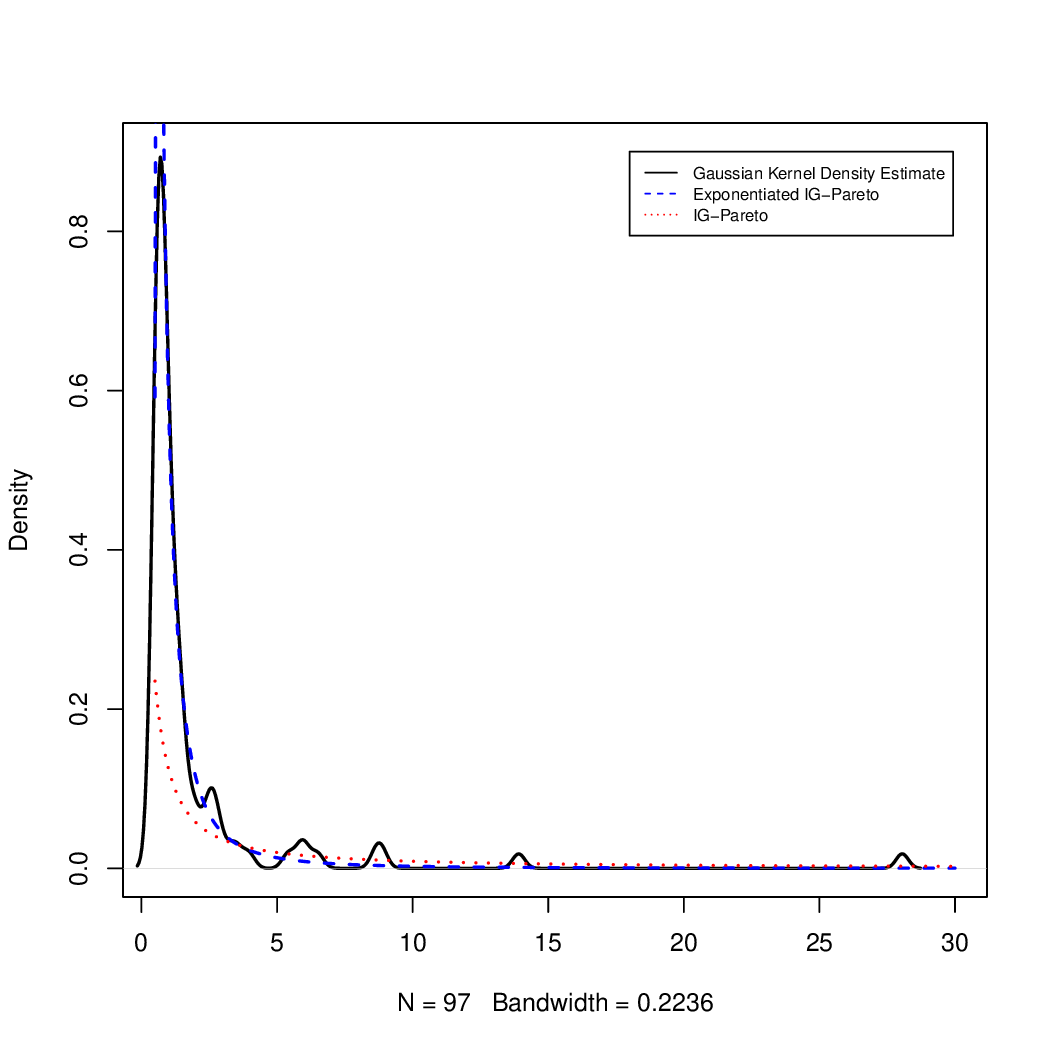}
	\caption{Density Plot of Nowrgewian Fire Insurance Data (year 1972) with corresponding exponentiated IG-Pareto and IG-Pareto model fit}
	\label{fig:densityplotnorway}
\end{figure}
\begin{figure}
	\centering
	\includegraphics[width=0.7\linewidth]{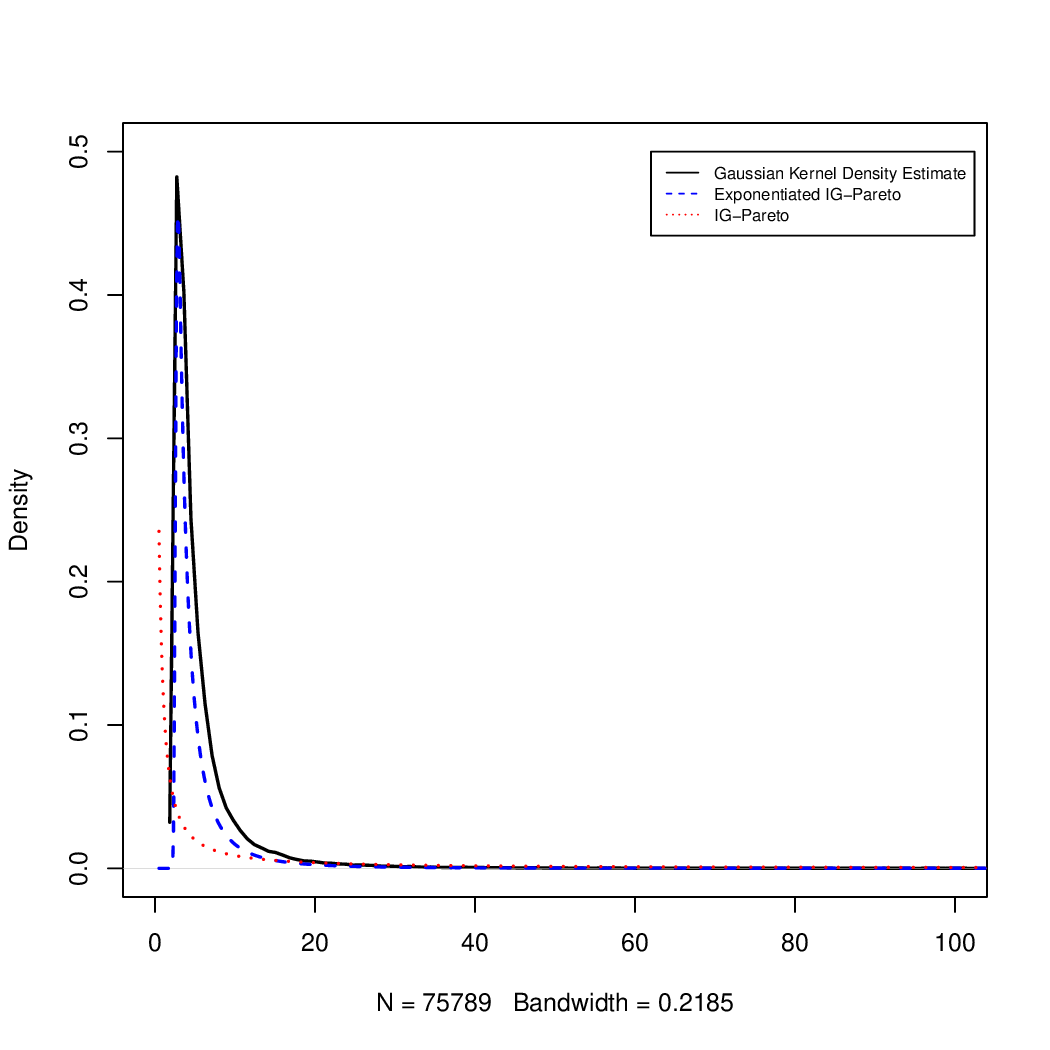}
	\caption{Density Plot of SOA Group Medical Large Claims with corresponding exponentiated IG-Pareto and IG-Pareto model fit}
	\label{fig:hist}
\end{figure}
\FloatBarrier
\section{Conclusion}
In this paper, we  proposed a new exponentiated IG-Pareto model to improve the performance of the original one-parameter IG-Pareto model. We provide an algorithm to find the MLE of $\theta$ and $\eta$ in Section 2. Such algorithm presents the ability to identify the MLE as the estimates for both $\theta$ and $\eta$ become more accurate as the sample size gets larger in all simulation scenarios. Three numerical examples are provided and the new exponentiated IG-Pareto model outperforms the original IG-Pareto model for all the examples. For the SOA Group Medical Insurance Large Claims data set, the exponentiated IG-Pareto model provided the best fit to the data among all the models.  The development of this model is promising since such exponentiation approach can also be applied to other composite models. 

\bibliography{ref}
\bibliographystyle{unsrt}  

\end{document}